\documentclass[12pt]{article} 
\usepackage{cite}
\usepackage{wrapfig}
\usepackage{amssymb}
\usepackage{amsfonts}
\usepackage{authblk}
\usepackage{amsmath}
\usepackage{lineno}
\usepackage{color}
\usepackage{xcolor}
\usepackage{multirow}
\usepackage{longtable}
\usepackage{lscape}
\usepackage[T2A]{fontenc}
\usepackage[utf8x]{inputenc}
\usepackage{hyperref}
\usepackage{bigstrut}

\begin{document}

\title{On the scale hierarchy in radiative symmetry breaking}

\author[1,2]{A.B. Arbuzov}
\author[2]{U.E. Voznaya}
\author[2]{T.V. Kopylova}

\affil[1]{\it Bogoliubov Laboratory of Theoretical Physics, 
                 Joint Institute for Nuclear Research, 
                 Joliot-Curie str. 6,  
                 141980 Dubna, Moscow region, Russia}
\affil[2]{\it Dubna State University,  
                 Universitetskaya str. 19, 141982 Dubna, Russia}

\date{}

\maketitle

\abstract{The emergence of a scale hierarchy in the case of spontaneous radiative breaking of conformal symmetry is discussed using the example of a simple quantum field theory model. The Coleman-Weinberg mechanism is implemented in the one-loop approximation of the effective potential of a scalar field interacting with a fermion field. The emergence of a hierarchy between the renormalization scale and the magnitude of the scalar field vacuum expectation value is shown. An effective model in the vicinity of the effective potential minimum is constructed, and absence of a direct renormalization group transition to the original theory is established. It is shown that effective model parameters measurement in the infrared region allows us to determine the scale that limits the applicability region of the model. \\[.3cm]
}

\section{Introduction}

Despite the extraordinary success of the Standard Model (SM) of particle physics
in the description of physical phenomena in all available for experimental research
energy ranges, there is no doubt that this model is not a true fundamental theory. 
Beyond the known complexities related to description of baryonic asymmetry, 
dark matter, and dark energy within the framework of the SM,
this model also contains a number of internal difficulties.
One of them is the so-called problem of naturalness or hierarchy. Indeed, direct calculations
in the framework of the SM and its extensions show that the energy scale in the Higgs boson sector
should be close to some {\em new physics} energy scale in practically any scenario. 
This circumstance served as a significant incentive for the search for new
physical phenomena at high energy colliders including LEP, Tevatron, and LHC.
The absence of any new physical phenomena manifestations up to the scale of the
1~TeV order makes the question of the SM energy scale origin being urgent.

The presence of Landau poles in the $U(1)$ and $\phi^4$ sectors of the SM also 
counts in favor of the fact that this model is an effective 
theory\footnote{In defining the concept of {\em effective theories}
we follow the work~\cite{Rivat:2019xrq}.}, 
the scope of which a priori has an upper bound of some large energy scale. 
Moreover, this scale turns out to be an additional parameter of the model 
in calculation of observables and, accordingly,
can be (should be) determined in a phenomenological analysis.

The energy scale of the Standard Model itself is introduced into it by setting a single
dimensionful parameter, namely, the tachyon mass of the primary scalar field.
The Brout--Englert--Higgs mechanism then generates masses of electroweak bosons
and the Higgs boson, which turn out to be of the same scale --- of the order of 100~GeV.
It should be noted that this mechanism is implemented
at the classical level and it is completely analogous to the Ginzburg--Landau approach 
to the description of superconductivity. The tachyon mass term introduced into the initial 
SM Lagrangian explicitly violates the scale invariance of the SM. This, in fact, leads to
the problem of naturalness, {\it i.e.}, to the question why the SM energy scale is so small
compared to the Planck mass or some not yet achieved experimentally scale of new physics. 
Note that in the framework of the effective theories it is more correct to raise 
the question of the relationship between the scale of the SM itself and the upper boundary 
of its applicability.

In addition, we note that, in contrast to the theory of superconductivity, a generally 
accepted microscopic description of the spontaneous symmetry breaking mechanism in 
the Standard Model is not yet found.
One of the most promising ways to construct such a mechanism is the assumption of
the fundamental nature of the conformal symmetry and implementation of spontaneous breaking
of this symmetry in one way or another. Indeed, conformal symmetry of
the quantum field model at the classical level, if spontaneously broken, can
naturally ensure the smallness of the generated masses and condensates
at the quantum level~\cite{Bardeen:1995kv}.
 
Within the framework of quantum field theory, spontaneous violation of conformal
invariance\footnote{Strictly speaking, we will consider
scale invariance violation, but at the same time
breaking of conformal invariance~\cite{Nakayama:2013is}.}
can be described in the Coleman-Weinberg (CW) approach~\cite{Coleman:1973jx}.
The CW mechanism allows one to describe the dimensional transmutation phenomenon 
in effective potentials of various models by considering radiative corrections.
The presence of conformal anomalies in such models forces one to introduce a
nonzero energy scale into the theory. The magnitude of this scale cannot 
be predicted due to the scale invariance of the original model at the classical level.
The arbitrariness of the choice of this scale is usually described using renormalization
group that is applicable specifically for scale invariant models.

However, in the effective theory approach, this scale takes on
a very specific physical meaning: it becomes the boundary of
the applicability region of the low-energy effective model~\cite{Rivat:2019xrq}.
In this paper, we discuss the physical meaning and the procedure of determination 
of the scale invariance violation parameter in the Coleman-Weinberg mechanism 
by the example of a simple model with one scalar self-interacting field of type 
$\varphi^4$ and one fermion field with Yukawa interaction.

\section{An example of the Coleman-Weinberg mechanism implementation}

Let us consider the Coleman-Weinberg mechanism using an example of a simple model with 
one scalar and one fermion field in the massless case. Let there be a quartic self-interaction
of the scalar field and Yukawa interaction between the scalar and the fermion fields.
The choice of this model is due to the following reasons.
Firstly, it is simple, but can be considered as a prototype of the
corresponding Standard Model sector\footnote{The gauge
interactions present in the SM are easy to include, but this will not affect the main 
topics discussed below.}. Secondly, adding the fermion sector
to a simple model like $\varphi^4$ is useful for expanding the perturbative region.
The Lagrangian of the model in question has the form
\begin{equation} \label{L_c}
\mathcal{L}_{c} = \frac{1}{2}(\partial_{\mu} \varphi_c)^2 - \frac{\lambda^2}{2}\varphi_c^4 
+ i \overline{\Psi}_c \gamma_\mu \partial_{\mu} \Psi_c - y \varphi_c \overline{\Psi}_c \Psi_c,
\end{equation}
where the index "$c$" \ indicates that we are in the (semi)classical approximation.
Note that this Lagrangian has the conformal symmetry, and mass
terms are forbidden for both fermion and scalar fields.

Let us consider quantum corrections in the one-loop approximation. We get
two contributions to the effective potential as corrections to the tree approximation
from scalar and fermion loops:
\begin{equation}
V_{\mathrm{eff}}=V_{\mathrm{tree}}+\Delta V_{s}+\Delta V_{f}.
\end{equation}
In the one-loop approximation, all graphs with one loop and with different number 
of vertices and external lines are summed.
For the scalar field, this term was obtained by Coleman and Weinberg~\cite{Coleman:1973jx}:
\begin{equation}
\Delta V_s = \frac{\lambda \Lambda^2}{64 \pi^2}\varphi_c^2 + \frac{\lambda^2 \varphi_c^4}{256 \pi^2} 
\left( \ln \frac{\lambda \varphi_c^2}{2 \Lambda^2} - \frac{1}{2} \right),
\end{equation}
where $\Lambda$ is the ultraviolet cut-off parameter of the loop integral.
Within the framework of the effective models ideology~\cite{Rivat:2019xrq} this parameter
sets the scale at which the chosen model~(\ref{L_c}) obviously stops working,
for example, due to the quantum gravity effects switching-on.

The contribution of the fermion loops is obtained in the similar way, it has the same structure:
\begin{eqnarray}
\Delta V_f = - \frac{N_c}{8 \pi^2} \biggl[ (y \varphi_c)^2 \Lambda^2
+ \frac{(y \varphi_c)^4}{2} \left( \ln{\frac{(y \varphi_c)^2}{\Lambda^2_f}} 
- \frac{1}{2} \right) \biggr],
\end{eqnarray}
where $N_c$ is the number of colors, $y$ is the Yukawa coupling constant.
According to the logic of effective model construction, the fermion loop cut-off parameter
$\Lambda$ should be selected equal to the one of the scalar loop,
although subsequent renormalization will lead to a result independent of this
choice.

After that we make renormalization following the procedure established in~\cite{Coleman:1973jx}.
It is important to note that the conformal symmetry embedded in Lagrangian~(\ref{L_c}) requires
introduction of a counter term. The latter has the form of a scalar field mass term
and completely cancels the obtained quadratically diverging terms.
In the SM, conformal invariance is initially explicitly broken. As a result it is impossible to
make such a cancellation without applying {\it ad hoc} fine-tuning conditions. 
This leads to the naturalness problem.

The logarithmic terms cannot be renormalized at
$\varphi_c = 0$ because of the infrared divergence. It forces us inevitably and independently
from the renormalization scheme to introduce a certain energy scale $M$, at which
the value of the coupling constant is normalized:
\begin{eqnarray}
\lambda = \frac{d^4 V(\varphi_c)}{d\varphi_c^4} \Bigg|_{\varphi_c = M}.
\end{eqnarray}
Note that the emergence of a new parameter in the model is inevitable
in any method of ultraviolet divergence regularization,
including dimensional regularization. However, the requirement of scale hierarchy
$M \ll \Lambda$ is essential in the procedure of effective
field theories construction, so a different from the original one effective theory arises on the scale $M$,
.
In our case, the renormalized effective potential contains a nonlocal term with 
a logarithmic field dependence:
\begin {equation} \label{Vren}
V_{\mathrm{\mathrm{eff}}}^{\mathrm{ren}} = \frac{\lambda \varphi_c^4}{4!} 
+ \frac{\varphi_c^4}{256 \pi ^2} \left[ \lambda^2 - 16 N_c y^4 \right] 
\left( \ln{\frac{\varphi_c^2}{M^2}} - \frac{25}{6} \right). 
\end{equation}

Let us now check where there are extremum points of the potential. We impose the
standard condition
\begin{equation}
\frac{dV_{\mathrm{eff}}^{\mathrm{ren}}(\varphi_c)}{d\varphi_c} = 0. 
\end{equation}
In our case, the minimum of the potential is shifted from the origin of the
coordinates to the point $\varphi_c = \langle \varphi_c \rangle \equiv v$, 
determined from the equation
\begin{equation} \label{dim_trans}
\ln\frac{\langle\varphi_c\rangle^2}{M^2} = \frac{11}{3} - \frac{\lambda 32\pi^2}{3[\lambda^2 - 16 N_c y^4]}. 
\end{equation}
The relations of such a type describe the so-called dimensional transmutation, 
that is possibility to express a dimensionless coupling constant (on the scale of
$M$) in terms of a dimensionful value $v$, which is absent in the initial Lagrangian.

If the scalar field is shifted onto its vacuum expectation value 
$\varphi_c = \varphi + v$, we are able to rewrite the nonlocal effective model 
obtained in the previous step in variables convenient for low-energy behavior analysis:
\begin{eqnarray} 
V_{\mathrm{eff}}^{\mathrm{ren}}(\varphi) = \frac{\lambda (\varphi+v)^4}{4!} 
+ \frac{(\varphi+v)^4}{256 \pi ^2} 
\left[ \lambda^2 - 16 N_c y^4 \right] 
\left( \ln{\frac{(\varphi+v)^2}{M^2}} - \frac{25}{6} \right). 
\end{eqnarray}
Following the logic of effective models construction, we expand the potential 
into a series near the minimum at small values of the field $\varphi \ll M$ 
(in the infrared region) and obtain
\begin{equation}
V_0(\varphi) =  \frac{m_0^2 \varphi^2}{2} + \frac{h_0 \varphi^3}{3!} 
+ \frac{\lambda_0 \varphi^4}{4!} 
+ \mathcal{O}\left(\varphi^4\frac{\varphi}{M}\right).
\end{equation}
The coupling constants and the mass of the scalar field in the resulting effective model are
\begin{eqnarray} 
\lambda_0 &=& \frac{\partial^4 V_{\mathrm{eff}}^{\mathrm{ren}}}{\partial \varphi^4} \Big|_{\varphi = 0} 
=\frac{11}{32 \pi^2} \left[ \lambda^2 - 16 N_c y^4 \right], \label{lambda0} \\
h_0 &=& \frac{\partial^3 V_{\mathrm{eff}}^{\mathrm{ren}}}{\partial \varphi^3} \Big|_{\varphi = 0} 
= \left[ \lambda^2 - 16 N_c y^4 \right] \frac{5 v}{32 \pi^2} = \lambda_0\frac{5 v}{11}, \label{h0} \\
m_0^2 &=& \frac{\partial^2 V_{\mathrm{eff}}^{\mathrm{ren}}}{\partial \varphi^2} \Big|_{\varphi = 0} 
= \frac{\left[ \lambda^2 - 16 N_c y^4 \right] v^2}{32 \pi^2} = \frac{\lambda_0 v^2}{11}.  \label{m0}
\end{eqnarray}
We stress that the coupling constant of the quartic interaction $\lambda_0$ in the arising
effective model is quadratic in the initial coupling constant $\lambda$. 
This means, in particular, that there is no continuous renormalization-group 
transition between the original model and the emerging one. 
So, the renormalization group evolution of $\lambda_0$ in the effective model 
does not lead to the correct high-energy asymptotics described by the original model. 
By construction this is because the range of applicability of an effective model with 
potential $V_0$ is limited to small values of the field $\varphi \ll M$.
This situation is typical for effective models.
Like in the SM, in our case the nonzero vacuum expectation value (VEV) of the scalar field 
in the Yukawa interaction generates the mass of the fermion field $m_f \sim yv$.

The generated triple self-coupling constant $h_0$ and masses $m_0$ and $m_f$ 
by construction are proportional to the VEV of the scalar field with coefficients 
of the order of unity, if there is no reason for extreme smallness or greatness of 
the initial coupling constants on the renormalization scale $M$.
So, all available for {\em measurement} dimensionful parameters of the low-energy 
effective model have one order of magnitude, which sets the energy scale of the model.

\section{Scale Hierarchy} 

Let us now discuss the relationship between the vacuum expectation value $v$ and the scale $M$.
In the approximation under consideration, it has the form
\begin{equation} \label{v}
v = M\exp{\left\{\frac{11}{6} - \frac{\lambda 16\pi^2}{3[\lambda^2 - 16 N_c y^4]} \right\}}, 
\end{equation}
{\it i.e.}, the proportionality coefficient may turn out to be
exponentially large or exponentially small depending on the values of the coupling constants.
In the work~\cite{Arbuzov:2015taa} it was shown for the case of simple
supersymmetric model that such a factor can separate $v$ and $M$ into many orders of magnitude.

In the original work of Coleman and Weinberg~\cite{Coleman:1973jx}
the renormalization scale $M$ was treated as an arbitrary value.
Based on this, it was proposed to choose it equal to $v$. In our opinion, such a choice
significantly violates the logic of effective models construction and, in fact,
is a big mistake. Indeed, for models with conformal anomaly, the scale arising 
in the CW mechanism has the character of an observable (measurable) quantity, 
albeit dependent on the scheme of its definition.
The assumption that the value of $M$ is arbitrary, essentially means
that the physical system retains the scale invariance, while the scale of $M$ was
introduced precisely as an invariance breaking parameter. A phenomenological example 
of such a situation is quantum chromodynamics: the conformal anomaly in it forces us 
to introduce the scale $\Lambda_{\mathrm{QCD}}$,
the specific meaning of which is determined from the analysis of phenomenological data.
In addition, imposing the condition $v = M$ in equation~(\ref{dim_trans}), 
we would obtain a relation linking dimensionless coupling constants. 
For the case of a single-interaction model, for example, $\sim \lambda \varphi^4$, 
we would {\em calculate} the value of the coupling constant
at the effective potential minimum point without even knowing the actual position of this minimum.
It is clear that this is due to the internal inconsistency of the condition $v = M$.

From a physical point of view, for the model with conformal symmetry on the classical 
level and conformal anomaly on the quantum level, the scale of the violation of scale invariance 
is determined by some external conditions, for example, boundary ones or a nontrivial
vacuum structure. 
When implementing the CW mechanism, $M$ was introduced for regularization
in the infrared region, although simultaneously ultraviolet divergences were renormalized. 
As shown above, the value of $M$ is not arbitrary, it is expressed in terms of the 
effective low-energy model parameters.

As noted above, the model~(\ref{L_c}) can be considered as a version of the Standard
Model reduced to conformal case and simplified by eliminating gauge interactions.
Let us estimate the scale of the hierarchy arising in our simplified case,
taking the known values of the Higgs boson VEV and mass, and the Yukawa coupling constant 
of the top quark. However, we do not claim to quantitative evaluation for realistic SM, 
where it is necessary to take into account the doublet structure of Higgs boson, 
interactions with vector bosons and known
higher order corrections to the effective potential\footnote{The corresponding analysis 
is in preparation.}. As the input for estimates, we take
\begin{equation}
m_0 = 125~\text{ГэВ}, \quad v = 246~\text{ГэВ}, 
\quad y=1, \quad N_c = 3.
\end{equation}
Substituting these values into~(\ref{lambda0}), (\ref{m0}), and (\ref{v}), we obtain
\begin{equation}
\lambda_0 \approx 2.8, \qquad
\lambda \approx 11.4, \qquad
M \approx 61~\text{ТэВ}.
\end{equation}
The found value of the scale limiting the applicability of the effective 
model shows the emergence of a hierarchy: $v \ll M$.

We also note that the constants $\lambda$ and $\lambda_0$ present in the estimates
satisfy the conditions of perturbativity at appropriate scales. Indeed, at the scale of 
$\varphi \sim M$ quantum corrections to the tree potential in~(\ref{Vren}) are small.
However, at the scale of $\varphi \sim v \ll M$, the contribution of the quantum 
corrections becomes of the order of the tree-level term. This means that the accuracy of 
the used one-loop approximation in estimation of the minimum position is unsatisfactory. 
For realistic estimates, higher-order contributions should be used, if possible, with their resummation.
However, the object of this paper was not the phenomenological evaluation, but the discussion 
of the stages of finding an effective model and of possibilities for an exponential scale 
hierarchy emergence.

\section{Conclusions}

This paper shows how the Coleman-Weinberg mechanism is implemented in the model
with one scalar and one fermion field taking one-loop corrections into account.
The question of choosing the renormalization scale $M$, which is also the
parameter of scale (at the same time conformal) invariance violation is considered.
We claim that it cannot be set equal to the vacuum expectation value $v$,
since for models with conformal anomaly the value of $M$ is
measurable, not arbitrary.
It is also shown that in the chosen model a hierarchy between
the scale of the conformal invariance violation and the generated masses of the scalar
and fermion fields naturally arises.
An effective theory that arises at small energy scales
and differs from the original theory is constructed.
In particular the absence of a renormalization group connection
between the theories is demonstrated.

The emergence of an exponential energy scale hierarchy was discussed
in a variety of models in condensed matter physics. So, for example, in the
Bardeen-Cooper-Schrieffer theory of superconductivity there is an exponential 
hierarchy between the value of the fermion condensate generated during spontaneous 
symmetry breaking (the Majorana mass) and the Debye frequency~\cite{Miransky:1993book}, 
which acts as an external condition for this system.
Also, there is an example is the recent work~\cite{Jian:2018aqk}, in which the emergence 
of the collective modes mass hierarchy in the infrared region near the quantum
critical point in the model describing high-temperature superconductors with pair 
density waves is discussed. In particular, in this work exponential smallness of 
the scalar mode mass (such as the Higgs boson) in comparison with the magnitude 
of the superconductor energy gap was established.

The authors are grateful to I.V.~Anikin and B.N.~Latosh for fruitful discussions 
and criticisms.

\end{document}